\documentclass [amssymb,amsmath,twocolumn, showpacs]{revtex4-1} 
\usepackage{graphicx,epsfig,amsfonts,amssymb}
\usepackage{bm}
\usepackage{times}
\usepackage{lipsum}

\newcommand{\beq}{\begin{equation}}
\newcommand{\eeq}{\end{equation}}
\newcommand{\bes}{\begin{subequations}}
\newcommand{\ees}{\end{subequations}}
\newcommand{\bea}{\begin{eqnarray}}
\newcommand{\eea}{\end{eqnarray}}
\newcommand{\ba}{\begin{array}}
\newcommand{\ea}{\end{array}}
\newcommand{\beqn}{\begin{eqnarray*}}
\newcommand{\eeqn}{\end{eqnarray*}}

\newcommand{\ra}{\rangle}

\newcommand{\upa}{\uparrow}
\newcommand{\dna}{\downarrow}

\begin{document}

\title{Solvable multistate model of Landau-Zener transitions in cavity QED}

\author {Nikolai~A. {Sinitsyn}$^{a}$ }
\author {Fuxiang {Li}$^{a,b}$ }
\address{$^a$ Theoretical Division, Los Alamos National Laboratory, Los Alamos, NM 87545,  USA}
\address{$^b$ Center for Nonlinear Studies, Los Alamos National Laboratory,  Los Alamos, NM 87545 USA}

\begin{abstract}
We consider the model of a single optical cavity mode interacting with  two-level systems (spins) driven by a linearly time-dependent  field. When this field passes through  values at which spin energy level splittings become comparable to spin coupling to the optical mode, a cascade of Landau-Zener (LZ) transitions leads to co-flips of spins in exchange for photons of the cavity.   
We derive exact transition probabilities between different diabatic states induced by such a sweep of the field.
\end{abstract}
\date{\today}

\maketitle

 

 
\section{Introduction}
Cavity quantum electrodynamics (cavity QED) is the study of  light interaction with discrete quantum systems, such as spins of electrons and  atoms, at conditions of significant coupling to a single photon.
Traditional realizations of cavity QED in atomic physics \cite{atomic-QED} have recently been extended to numerous solid state applications, including superconductors \cite{supercond-QED}, spins of defects and quantum dots \cite{QD-QED} and fiber optics \cite{roy-QED}. 
One of the most fundamental models in cavity QED is the Tavis-Cummings model \cite{tv} that describes coupling of $N_s$ two-level systems, which we will call spins for simplicity, to a single optical cavity mode  and an external field. The Hamiltonian of this model is given by \cite{note-qed}
\beq
\hat{H}=\omega \hat{a}^{\dagger} \hat{a}- \Delta \sum_{i=1}^{N_s} \hat{\sigma_i}  +\sum_{i=1}^{N_s} \epsilon_i \hat{\sigma}_i +g\sum_{i=1}^{N_s} (\hat{a}^{\dagger}\hat{\sigma}_i^- + \hat{a} \hat{\sigma}_i^+ ),
\label{ham1}
\eeq 
where $\omega$ and $\hat{a}$ are, respectively, the frequency and the annihilation operator of the cavity mode,  and $\hat{\sigma}_i^{\pm}$ are $i$th spin's raising/lowering matrices.
Parameter $g$ describes  spin couplings to the optical mode. This coupling is assumed to be identical for all spins, as in the original Tavis-Cummings model \cite{tv}; $\Delta$ is the Zeeman-like splitting induced by the external  field.  We include the possibility of spins to experience additional local level splittings described by arbitrary parameters $\epsilon_i$. 
We also introduced projection operators 
$$
\hat{\sigma}_i \equiv (\hat{1}_i +\hat{\sigma}_z^i)/2,
$$
to spin ``up" states. Here $\hat{1}_i$ is a unit matrix acting in the $i$th spin subspace, and $\hat{\sigma}_z^i$ is the Pauli z-matrix of the $i$th spin. 
This model can be physically derived from a more general  Dicke model \cite{dicke} by applying the rotating wave approximation. The latter is well-justified in numerous available realizations of the cavity QED.

The model (\ref{ham1}) conserves the number of excitations,  i.e. the Hamiltonian commutes with the operator  of the number of bosons plus the number of spins up:
\beq
\hat{N}_e= \hat{a}^{\dagger} \hat{a}+\sum_{i=1}^{N_s} \hat{\sigma}_i.
\label{inv}
\eeq
Therefore, it is usually possible to diagonalize the matrix Hamiltonian numerically for small $N_s$ and  $N_e$. However, the size of the phase space grows quickly with $N_e$ and $N_s$, so further approximations are usually invoked, such as the degeneracy assumption that all parameters $\epsilon_i$ are identical. Various limits and extensions of the model (\ref{ham1}) at constant parameters have been solved with algebraic Bethe ansatz \cite{tv-bethe}. 

Cavity QED is usually discussed in the context of achieving control of  quantum states of two level systems and the photon mode. For example, practical goals can be to implement gates for quantum computation with spin qubits or
 to use spins as emitters of strongly nonclassical photon states.  Such a control requires the ability to change parameters of the system in time. Therefore, time-dependent versions of the cavity QED models currently garner considerable scientific attention \cite{cavity-QED-time,cavity-QED-control}.  
 
 Landau-Zener (LZ) transitions \cite{book-LZ} induced by a linear sweep of parameters in cavity QED models have been suggested as a tool to produce a strongly non-classical light source and entangle qubits in an optical cavity  \cite{cavity-QED-LZ, cavity-QED-MLZ}. In the case of the model (\ref{ham1}), this corresponds to the assumption that 
 \beq
 \Delta =\beta t,
 \label{time1}
 \eeq
 with some constant sweeping rate $\beta$ and $t$ is time.  Beyond cavity QED, similar models of linearly driven interacting spin systems have been recently discussed for applications in ultracold atomic gases \cite{atomic,app-bose}, LZ-interferometry \cite{LZ-interferometry}, quantum dots \cite{dot-lz-exp1}, quantum control \cite{qcontrol},  and quantum coherence \cite{coher}.    
 
 Considering Eq.~(\ref{time1}) at $t\rightarrow \pm \infty$, one can disregard the last sum  in the Hamiltonian (\ref{ham1}) so that its eigenstates  become simultaneously the eigenstates of the photon number operator, $\hat{a}^{\dagger} \hat{a}$,  and spin polarization operators, $\hat{\sigma}_z^i$.  
 We will call such states the {\it diabatic states}.   Imagine then that at large negative time $t$, the system is in the state without photons in the cavity mode and with all spins polarized in the diabatic state
  $|0;\upa_1 \upa_2 \ldots   \upa_{N_s} \ra$. As time changes, eventually, spin splittings become comparable to the coupling strength $g$ so that some of the spins flip by emitting photons. Since photons are bosons, the increase of their number 
  enhances the coupling of the photon mode to all the spins, which facilitates new transitions and so on.  
   Such nonlinear effects can be used in practice to induce strongly correlated spin-photon states. 
   
    Importantly, at large positive time $t$ all spin levels again split out of the resonance so that all transitions terminate.  There is  a characteristic time $\sim g/\beta$, during which transitions between diabatic states are essential. We will assume 
    sufficiently high quality optical cavity, for which this time is much shorter than the time of photon loss, so we can disregard dissipation. One can define then the following scattering problem: Given Eqs.~(\ref{ham1}), (\ref{time1}) and  that at $t\rightarrow -\infty$
   the system is in one of the diabatic states of the Hamiltonian $\hat{H}$,  marked by an index $n$, what is the  probability $P_{n'n} \equiv P_{n\rightarrow n'}$ to find the system in a diabatic state with an index $n'$ at 
$t\rightarrow +\infty$? We will call such a scattering problem exactly solvable if there is a procedure to write down any element $P_{n'n}$  of the {\it transition probability matrix} $\hat{P}$ explicitly in terms of the well known special functions of model parameters.

The time-dependent problem with the Hamiltonian (\ref{ham1}) under assumption (\ref{time1}) looks very complex at first view. In order to find scattering probabilities between microstates one has to solve a system of many coupled differential equations with linearly changing time-dependent coefficients. Solutions of such equations show violant  oscillations, which decay with long power-law tails in time. In addition, the number of coupled equations is growing exponentially with $N_s$. Therefore, generally, such systems remain challenging to study even numerically at $N_s>4$.   

In this article we will argue, however, that the scattering problem for this model is exactly solvable, i.e., we will provide an iterative procedure to obtain transition probabilities between any pair of states in any sector of the model in a finite number of steps. We will also investigate some of the consequences of this solution.  

The plan of our article is as follows. In section 2, we will describe the connection of the model (\ref{ham1})-(\ref{time1}) to the multistate LZ problem and explore simplest situations  that reduce to the previously solved models. In section 3, we will present solution of the model (\ref{ham2}) in the eight-state sector, which corresponds to one of the simplest nontrivial cases of the Hamiltonian (\ref{ham1}) that do not reduce to the already known solvable multistate LZ models. 
We discuss higher dimensional sectors in section 4.
In section 5, we consider the degenerate case $\epsilon_i=0$ for all $i$ and derive the full transition probability matrix for the four-state sector, and investigate transition probabilities from the fully polarized initial conditions at arbitrary size of the phase space. In section 6, we will summarize our major findings and discuss perspectives to explore other strongly driven explicitly time-dependent interacting quantum problems.

\section{Driven Tavis-Cummings model as a multistate Landau-Zener problem}

We will start our discussion with slightly rewriting the Hamiltonian in order to reduce the number of independent parameters. Due to conservation of the number of excitations (\ref{inv}), we can replace the operator $\sum_{i=1}^{N_s} \hat{\sigma_i}$ with $N_e-\hat{a}^{\dagger} \hat{a}$ in any sector with a constant $N_e$.
The term with $N_e$ commutes then with the rest of the Hamiltonian and can be safely removed by a gauge transformation.
Hence, we can rewrite the fist two terms in (\ref{ham1}) as  $(\omega +\beta t)\hat{a}^{\dagger} \hat{a}$. If only the evolution from infinitely large negative to infinitely large positive time values is considered, the redefinition of time, $t \rightarrow t-\omega/\beta$, does not affect boundary conditions and, consequently, the scattering probabilities. On the other hand, it removes parameter $\omega$ from the Hamiltonian. 
Moreover, rescaling time as $t \rightarrow t/\sqrt{ \beta}$ in the Schr\"odinger equation, $i\dot{\psi}(t)=\hat{H}(t) \psi(t)$, where $\psi(t)$ is the state vector,
and redefining couplings as $g/\sqrt{\beta} \rightarrow g $ and $ \epsilon_i/\sqrt{\beta}  \rightarrow \epsilon_i$, parameter $\beta$ drops out of this equation. Therefore, the Hamiltonian of the problem simplifies:
\beq
\hat{H}= t \hat{a}^{\dagger} \hat{a}  +\sum_{i=1}^{N_s} \epsilon_i \hat{\sigma}_i +g\sum_{i=1}^{N_s} (\hat{a}^{\dagger}\hat{\sigma}_i^- + \hat{a} \hat{\sigma}_i^+ ).
\label{ham2}
\eeq 
Let us first consider $N_e=1$. In this case,  the available phase space consists of the state with a single photon:
\beq
|N_s+1\ra \equiv |1;\dna_1, \dna_2, \ldots \dna_{N_s} \ra,
\label{diab1}
\eeq
and $N_s$ states with a single spin excitation:
\beq
 \quad |i\ra \equiv |0;\dna_1, \dna_2, \ldots ,\uparrow_i, \ldots \dna_{N_s} \ra.
\label{diab2}
\eeq

In order to enumerate different diabatic states, here and in what follows we will assume the convention in which the state with the minimal possible number of photons has index 1; diabatic states with the same number of photons are labeled in the decreasing order of their energy at $t \rightarrow - \infty$ (when disregarding coupling terms in the Hamiltonian), and states with a larger number of photons have higher indexes than the states with a smaller number of photons. Let $\epsilon_1>\epsilon_2> \ldots > \epsilon_N$. 
Following this convention, the Hamiltonian (\ref{ham2}) in the basis (\ref{diab1})-(\ref{diab2}) has the matrix form:
\begin{equation}
\hat{H}=\left( 
\begin{array}{ccccc}
\epsilon_1   & 0                   &\ldots     & 0                  &g  \\
0                  & \epsilon_2  & 0           & \ldots           &g  \\
\vdots                   & \vdots                 &   \ddots &\vdots           & \vdots \\
0        &\ldots         &  0	         &\epsilon_{N_s}    & g\\
g                   & g                  & \ldots   & g         &t
\end{array}
\right).
\label{do-ham}
\end{equation} 

At this point, we note that the Hamiltonian (\ref{do-ham}) and the more general Hamiltonian (\ref{ham2}) belong to the class of, so-called, multistate LZ models that describe the evolution of a  number $N$ of  states according to the Sch\"odinger equation with parameters that change linearly with time \cite{be}:
\begin{equation}
i\frac{d\psi}{d t} = \hat{H}(t)\psi, \quad \hat{H}(t) = \hat{A} +\hat{B}t .
\label{mlz}
\end{equation} 
Here, $\psi$ is the state vector in a space of $N$ states; $\hat{A}$ and $\hat{B}$ are constant Hermitian $N\times N$ matrices.  One can always choose the, so-called, {\it diabatic basis} in which the matrix $\hat{B}$ is diagonal,
and if any pair of its elements are degenerate then the corresponding off-diagonal element of the matrix $\hat{A}$ can be set to zero by a time-independent change of the basis, that is
\beq
B_{ij}= \delta_{ij}\beta_i, \quad  A_{nm}=0\,\,\, {\rm if} \,\, \beta_{n}=\beta_{m},\,\,n\ne m \in (1,\ldots N).
\label{diab3}
\eeq
Constant parameters $\beta_i$  are called the {\it slopes of diabatic levels},  nonzero off-diagonal elements of the matrix $\hat{A}$ in the diabatic basis are called the {\it coupling constants},  and the diagonal elements of the Hamiltonian
$$
H_{ii}=\beta_i t +\varepsilon_i,
$$
 where $\varepsilon_i \equiv A_{ii}$, are called the {\it diabatic energies}.
 Some of the multistate LZ models have been already studied  with the purpose to obtain transition probabilities between diabatic states \cite{do,bow-tie,bow-tie1,multiparticle,no-go,mlz-1,six-LZ,four-LZ}. For example, the model (\ref{do-ham}) is a special case of
  the exactly solvable Demkov-Osherov model  \cite{do}. 
  
\begin{figure}
\scalebox{0.10}[0.10]{\includegraphics{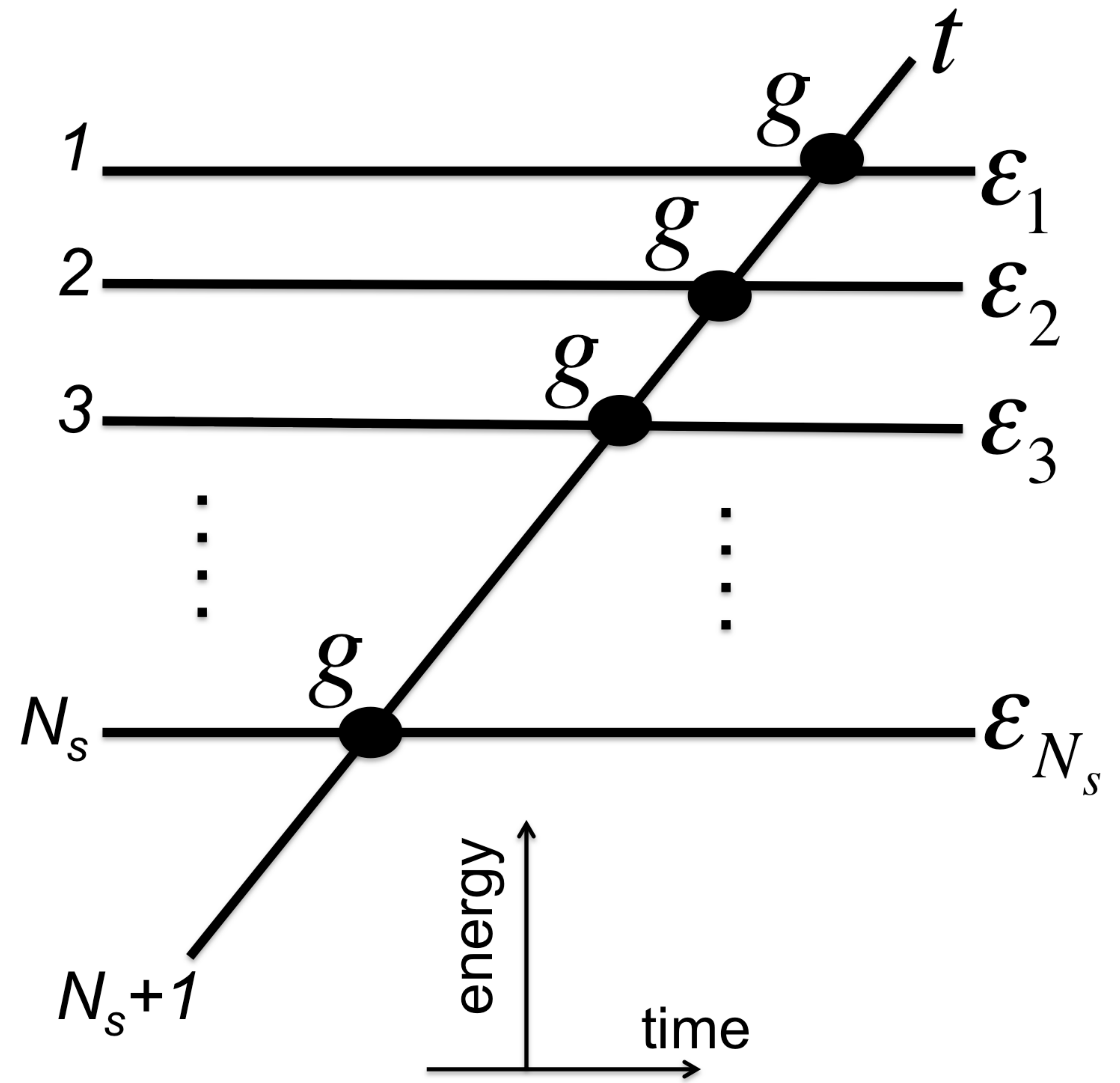}}
\hspace{-2mm}\vspace{-4mm}   
\caption{ Diabatic levels and couplings between them described by the Hamiltonian (\ref{ham2}) at $N_e=1$.}
\label{do-levels}
\end{figure}

  It is convenient to represent any multistate LZ-model in the time-energy diagram that plots diabatic energies with intersections of diabatic energy levels marked by corresponding pairwise coupling constants, as we show in Fig.~\ref{do-levels} for the model (\ref{do-ham}). According to this figure, the Demkov-Osherov model describes the case of a single  level intersecting a band of parallel diabatic levels.

It turns out that, irrespectively of the coupling strength, transition probabilities in the Demkov-Osherov model can be obtained following a simple rule: One should obtain a semiclassical trajectory that connects the initial and final states in the graph in Fig.~\ref{do-levels} and assume that probabilities to pass, $p$, or turn, $q$, at any level intersection are described by the standard LZ formula for intersection of two levels:
\beq
p=e^{-2\pi g^2}, \quad q=1-p.
\label{lz1}
\eeq 
Only semiclassical trajectories that turn or pass forward in time are allowed, so there is, maximum, a single trajectory that connects any given pair of states for the Demkov-Osherov model.   
The probability of the full trajectory is then the product of all  encountered pairwise passing or turning probabilities. If there is no such a semiclassical trajectory that connects two given diabatic states, the corresponding transition probability is zero.
 For example, if evolution starts with the state $|N_s+1\ra$ with a single photon, then the probability to remain in this diabatic state at $t\rightarrow +\infty$ is given by 
\beq
P_{N_s+1,N_s+1}= \prod_{i=1}^{N_s} p =p^{N_s}.
\eeq
Similarly, e.g, 
\beq
P_{21}= 0, \quad P_{22}=P_{11}=p, \quad P_{13}=q^2p.
\eeq
We remind that we use the convention in which $P_{21}$ means the probability of the transition {\it from} level-1 {\it to} level-2.

\begin{figure}
\scalebox{0.10}[0.10]{\includegraphics{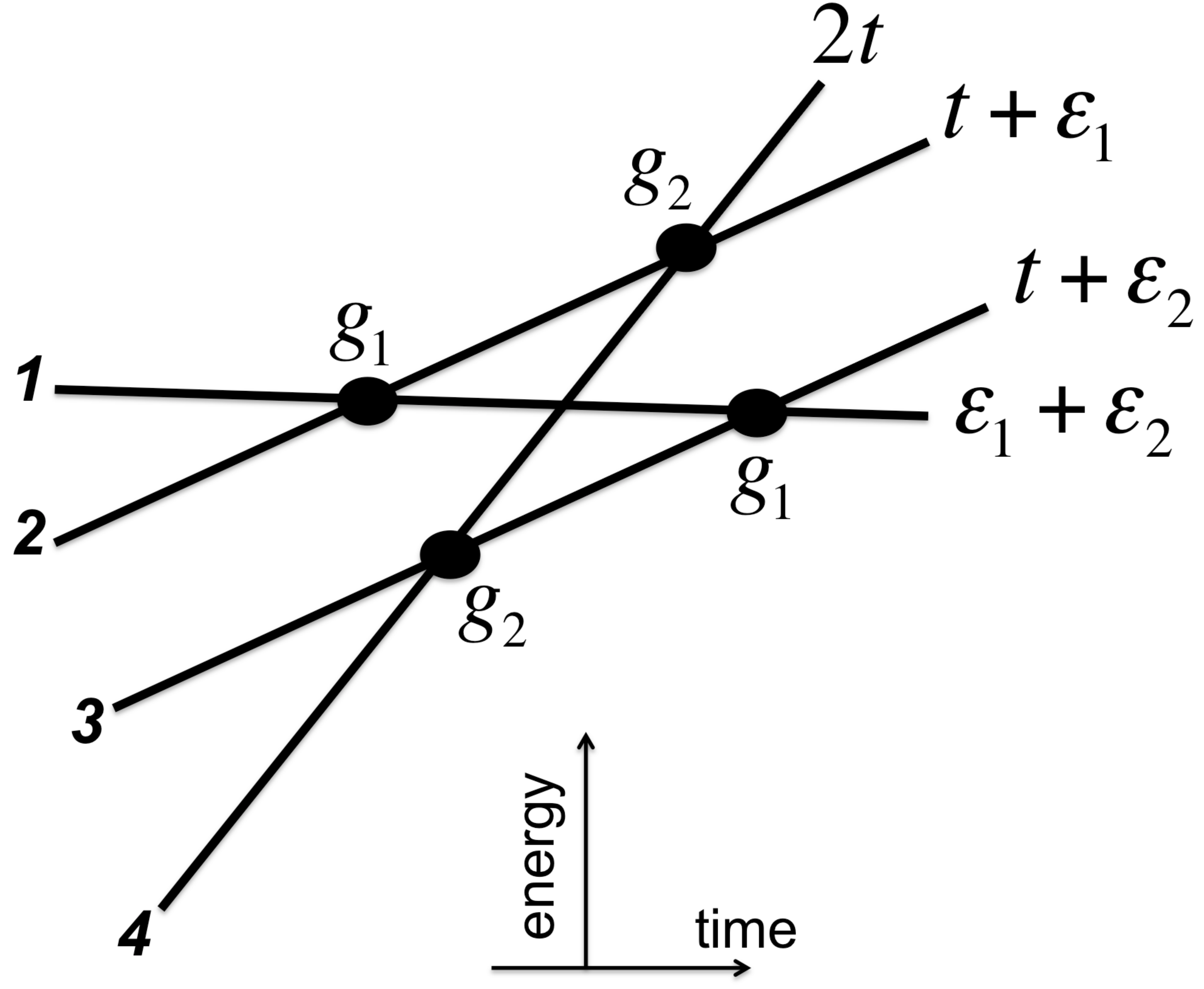}}
\hspace{-2mm}\vspace{-4mm}   
\caption{  Diabatic levels and couplings between them described by the Hamiltonian (\ref{bow-tie}) at $N_B=0$.}
\label{bow-tie-fig}
\end{figure}

Consider now another case that reduces to another multistate LZ-model, which solution is known. Assume that we have an arbitrary minimal number of bosons $N_B$ but only two spins. The diabatic basis consists then of only four states: 
\begin{eqnarray}
\label{states-4}
|1\ra &=& |N_B\upa_1 \upa_2 \ra,  \quad |2\ra = |N_B+1;\upa_1 \dna_2 \ra, \\
\nonumber |3\ra & =& |N_B+1;\dna_1 \upa_2 \ra, \quad |4\ra = |N_B+2;\dna_1 \dna_2 \ra.
\end{eqnarray}
Let us denote $g_1 = g\sqrt{N_B+1}$, and $g_2=g\sqrt{N_B+2}$. The Hamiltonian (\ref{ham2}) in the diabatic basis then reads:  
\begin{widetext}
\begin{equation}
\hat{H}=\left( 
\begin{array}{cccc}
 \epsilon_1+\epsilon_2 +N_B t& g_1                                 &g_1                                     & 0              \\
g_1                                                     & \epsilon_1+(N_B+1) t  & 0                                         & g_2         \\
g_1                                                     & 0                                      & \epsilon_2+ (N_B+1) t    & g_2          \\
0                                                          & g_2                                 & g_2                                    &  (N_B+2) t        
\end{array}
\right).
\label{bow-tie}
\end{equation} 
\end{widetext}
In Fig.~\ref{bow-tie-fig}, we show diabatic levels of this model. Although slopes of the levels are plotted there for the case $N_B=0$, we note that arbitrary $N_B$ would merely increase the slopes of all levels by equal amount, keeping relative slopes between levels the same, so geometry and the chronological order of level crossings would not change.  
The model in Fig.~\ref{bow-tie-fig} is recognized as a four-state  generalized bow-tie model, which exact solution is known \cite{bow-tie}. The generalized bow-tie model generally describes the case when several diabatic levels interact with two other parallel levels but do not interact with each other directly. Each of the levels from the first set couples equally to the parallel levels. Moreover, all diabatic levels from the first set must intersect at a single point - right in the middle between the parallel levels. It is easy to check that those conditions are satisfied by levels of the Hamiltonian (\ref{bow-tie}).

It has been noticed by Demkov and Ostrovsky  that the semiclassical procedure that we described above for the Demkov-Osherov model can be equally applied to determine {\it exact} solution of the bow-tie model \cite{bow-tie}: As in the case of the Demkov-Osherov model, there is a unique path in Fig.~\ref{bow-tie-fig} that preserves causality and connects any given pair of diabatic states at $t \rightarrow \pm \infty$. Let us denote
$$
p_1=e^{-2\pi g^2(N_B+1)}, \quad p_2=e^{-2 \pi g^2 (N_B+2)},
$$
and $\quad q_i =1-p_i$; $i=1,2$. Then the exact form of the transition probability matrix for the model (\ref{bow-tie}) can be readily written using Fig.~\ref{bow-tie-fig} and semiclassical rules:
\begin{equation}
\hat{P}=\left( 
\begin{array}{cccc}
 p_1^2     & q_1 p_1           &p_2q_1     & q_1q_2	 \\
q_1p_2      & p_1 p_2           & q_2^2     &  p_2q_2     \\
p_1q_1      & q_1^2            & p_1p_2    & p_1q_2         \\
q_1q_2      & p_1q_2            & p_2q_2     &  p_2^2        
\end{array}
\right).
\label{bow-tie-prob}
\end{equation}

One more important observation made by Demkov and Ostrovsky in \cite{bow-tie} is that the degenerate case of a generalized bow-tie model, which in our case corresponds to $\epsilon_1=\epsilon_2=0$, reduces to a model, which exact solution can be trivially obtained from the knowledge of the transition probability matrix of the nondegenerate case, such as (\ref{bow-tie-prob}). Let $ a(t),b_1(t),b_2(t),c(t) $ be the amplitudes of states, respectively, $ |1\ra, |2\ra, |3 \ra, |4\ra $ defined in (\ref{states-4}). In the degenerate case, i.e., at $\epsilon_1=\epsilon_2=0$, the Schr\"odinger equation for these amplitudes with the Hamiltonian (\ref{bow-tie})  corresponds to a set of four coupled differential equations:
\begin{eqnarray}
\label{eq1b}
\nonumber i\dot{a}&=&N_Bt a + g_1 (b_1+b_2), \\
\nonumber i\dot{b}_1 &=&(N_B+1)t b_1 +g_1a +g_2 c,\\
\nonumber i\dot{b}_2 &=&(N_B+1)t b_2 +g_1a +g_2 c,\\
i\dot{c} &=&(N_B+2)t c +g_2(b_1+b_2).
\end{eqnarray}
Let us now introduce new variables $b_{\pm}=(b_1\pm b_2)/\sqrt{2}$. Then the amplitude $b_{-}$ completely decouples while the remaining amplitudes satisfy a new differential equation:
\begin{equation}
i\frac{d}{dt}\left( \begin{array}{l}
a\\
b_+\\
c
\end{array} \right)=
\hat{H}^{(3)} \left( \begin{array}{l}
a\\
b_+\\
c
\end{array} \right),
\label{bow-tie-short}
\end{equation} 
where
\beq
\hat{H}^{(3)}=\left( 
\begin{array}{ccc}
N_Bt & g_1 \sqrt{2} &0 \\
g_1\sqrt{2} & (N_B+1)t & g_2 \sqrt{2} \\
0 & g_2\sqrt{2} & (N_B+2)t
\end{array}
\right).
\label{three-LZ}
\eeq
Equation~(\ref{bow-tie-short}) corresponds to a new three-state LZ model in which all diabatic levels intersect in one point at $t=0$.  This model is also exactly solvable, and it is known as a standard bow-tie model \cite{bow-tie1}. Comparing exact solutions for standard and generalized bow-tie models, Demkov and Ostrovsky  pointed that transition probabilities between states $|1\ra$ and $|3\ra$ of the model (\ref{three-LZ}) are the same as transition probabilities between states 
$|1\ra$ and $|4\ra$  in the nondegenerate version of the four-state model (\ref{bow-tie}),  while transition probabilities from or to the level $|2\ra$ of the model (\ref{three-LZ}) are given  by summation of transition probabilities, respectively, from and to both parallel levels with amplitudes $b_1,b_2$ in the nondegenerate version of the model (\ref{bow-tie}). For example,
$$
P_{12}^{(3)} =P_{12}+P_{13}, \quad P_{23}^{(3)}=P_{24}+P_{34},
$$
where the index ``(3)" marks transition probabilities in the model (\ref{bow-tie-short}), and probabilities $P_{ij}$ are defined in (\ref{bow-tie-prob}). Hence, the full transition probability matrix for the model (\ref{bow-tie-short}) reads: 
\begin{equation}
\hat{P}^{(3)}=\left( 
\begin{array}{ccc}
 p_1^2     & q_1(p_1+p_2)           &q_1q_2     	 \\
q_1(p_1+p_2)      &(1- p_1- p_2)^2           & q_2(p_1+p_2)       \\
q_1q_2     & q_2(p_1+p_2)          & p_2^2            
\end{array}
\right),
\label{bow-tie-prob-short}
\end{equation}
where the element $P^{(3)}_{22}$ was determined merely from the fact that, for unitary evolution, the transition probability matrix must be doubly stochastic, i.e. the sum of its elements along any column or any raw should be equal to one. Here, we make a new observation that 
\beq
1+P^{(3)}_{22}=P_{22}+P_{33}+P_{23}+P_{32},
\eeq
where, on the right hand side, we have the sum of all transition probabilities between parallel levels in the four-state model (\ref{bow-tie}), and the left hand side
is the sum of the probability to stay at the central level in the model (\ref{bow-tie-short}) plus 1. This unit, 1, can be interpreted as the probability, $P_{b_{-} \rightarrow b_{-}}$, to 
stay in the antisymmetric state with amplitude $b_{-}$ if the evolution starts with $b_{-}=1$.    We will use this observation to derive transition probabilities in the higher dimensional sectors of the degenerate version of the model (\ref{ham2}).

 \section{Eight dimensional sector}
  \subsection{Parameters of the Model}
Let us first introduce the notation to be used in this section and later on:
\beq
g_n=g\sqrt{N_B+n}, \quad p_n=e^{-2\pi g^2(N_B+n)}, \quad q_n=1-p_n,
\label{ppp}
\eeq
where  $n$ is a positive integer, and $N_B$ is the minimal number of photons that can be in the given sector of the model, i.e. $N_B=N_e-N_s$ for $N_e \ge N_s$. Formally, $N_B$ is an integer parameter of the model (\ref{ham2}), however, our subsequent discussion will apply equally to continuation of this parameter to any positive real value.  In this sense, our following discussion goes beyond the scope of the model (\ref{ham1}). We also note that negative integer values of $N_B$ are also allowed as far as $N_e>0$. In such a case, only the sector of states coupled by positive real couplings (\ref{ppp}) should be considered physical. We will focus only on sectors with $N_e\ge N_s$, although treatment of cases with $N_e<N_s$ is  similar. 
We will also introduce two parameters:
\beq
 N\equiv 2^{N_s},
\label{pars1}
\eeq
and a nonnegative integer parameter $M(n)$ that is the  number of spins ``down"  in the state number $n$, e.g., $M(1)=0$, $M(2)=1$, $M(N)=N_s$, e.t.c..
The meaning of $N$ is the number of independent connected states in the sector with $N_s$ spins-1/2 when $N_e \ge N_s$.

\subsection{Diabatic and Adiabatic Levels}

The nearest sector of the model (\ref{ham2}) with more than two spins has $N_s=3$, which, for $N_e\ge N_s$, corresponds to dimensionality of the model $N=8$ with the Hamiltonian:
\begin{widetext}
\beq
\hat{H}=\left( 
\begin{array}{cccccccc}
\epsilon_1+\epsilon_2+\epsilon_3   & g_1 &g_1 & g_1 &0 & 0  &0 & 0 \\
g_1 & \epsilon_1+\epsilon_2+t & 0 & 0 & g_2 & g_2 & 0 & 0 \\
g_1& 0 & \epsilon_1 +\epsilon_3 +t & 0 & g_2 & 0 & g_2 & 0 \\
g_1 & 0& 0& \epsilon_2+\epsilon_3 +t & 0 & g_2 & g_2 & 0 \\
0 & g_2 & g_2 & 0 & \epsilon_1 +2t & 0 & 0 & g_3 \\
 0 & g_2 & 0 & g_2 & 0 & \epsilon_2 +2t & 0 & g_3 \\
0 & 0 & g_2 & g_2 & 0& 0 & \epsilon_3 +2t & g_3\\
0 & 0 & 0 & 0 & g_3 & g_3& g_3 & 3t 
\end{array}
\right).
\label{ham-8}
\eeq
\end{widetext}  
Here, for simplicity, we dropped the term $N_B t \hat{1}$ from the main diagonal of the matrix (\ref{ham-8}) because this term can be removed by a time-dependent change of phases of all diabatic state amplitudes \cite{note-qed,be}; therefore  it does not influence transition probabilities and can be disregarded. Note that the definition of couplings (\ref{ppp}) still depends on $N_B$.

Our major method to study  models like (\ref{ham-8}) is based on the observation   that, surprisingly, all known exactly solvable models of the type (\ref{mlz}) with a finite number of interacting states  \cite{multiparticle,do,be,bow-tie,six-LZ,four-LZ} have exact solutions for the scattering matrix that coincide with  predictions of the independent crossing approximation, which we  described for cases of Demkov-Osherov and bow-tie models in previous section. Moreover, all such solvable models have two properties \cite{six-LZ,four-LZ}: 

 (i) the absence of the dynamic phase effect on transition probabilities in the semiclassical framework, and 
 
 (ii) the existence of exact crossings of eigenenergy levels (i.e. instantaneous eigenvalues of the Hamiltonian as functions of time) that correspond to each intersection of diabatic levels (i.e. diagonal elements of the Hamiltonian in the diabatic basis) without direct couplings in the diabatic level diagram. 

Here by the dynamic phase we mean the trivial phase 
\beq
\phi_{\rm dyn} = -\int_{-\infty}^{\infty} [ \beta_{k(t)} t +\varepsilon_{k(t)}] \, dt,  
\label{dynph}
\eeq
where $k(t)$ is the index of the level along a semiclassical trajectory. Its time-dependence follows from the possibility of switching   diabatic level indexes at level crossing points. 

{\it LZ-integrability conditions} (i)-(ii) have already been used by one of us to search for new solvable multistate LZ systems. There are already three different such models that have been found and solved in \cite{six-LZ,four-LZ} despite rigorous mathematical proof for their  integrability is still missing. The major observation of the present article is that invariant sectors of the model (\ref{ham2}), such as the $8\times 8$ model (\ref{ham-8}), in the nondegenerate case (e.g., $\epsilon_1>\epsilon_2>\epsilon_3$), also satisfy conditions of integrability (i)-(ii); therefore, corresponding matrices of transition probabilities can be obtained exactly by applying the independent crossing ``approximation".   
\begin{figure}
\scalebox{0.10}[0.10]{\includegraphics{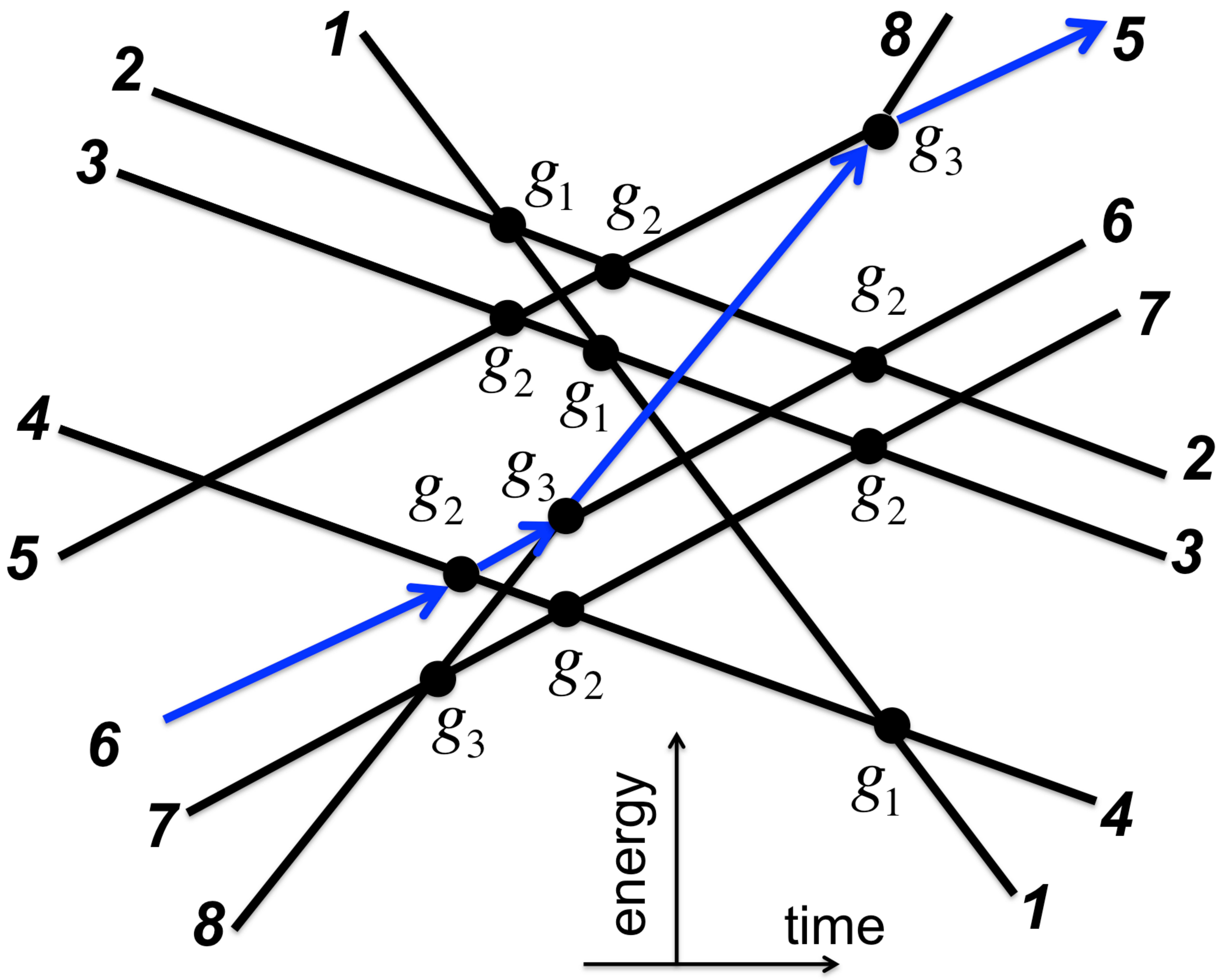}}
\hspace{-2mm}\vspace{-4mm}   
\caption{ Diabatic levels and couplings between them described by the Hamiltonian (\ref{ham-8}). For convenience, equal terms are added to all slopes of diabatic levels: $\hat{H}\rightarrow \hat{H}-1.5t \hat{1}$. This is equivalent to a change of time-dependent phases of state amplitudes, which  does not change transition probabilities. Blue arrows demonstrate the unique semiclassical path connecting the state $|6\ra$ at $t\rightarrow - \infty$ to the state $|5\ra$ at $t\rightarrow +\infty$.}
\label{fig-h8}
\end{figure}

\begin{figure}
\scalebox{0.34}[0.34]{\includegraphics{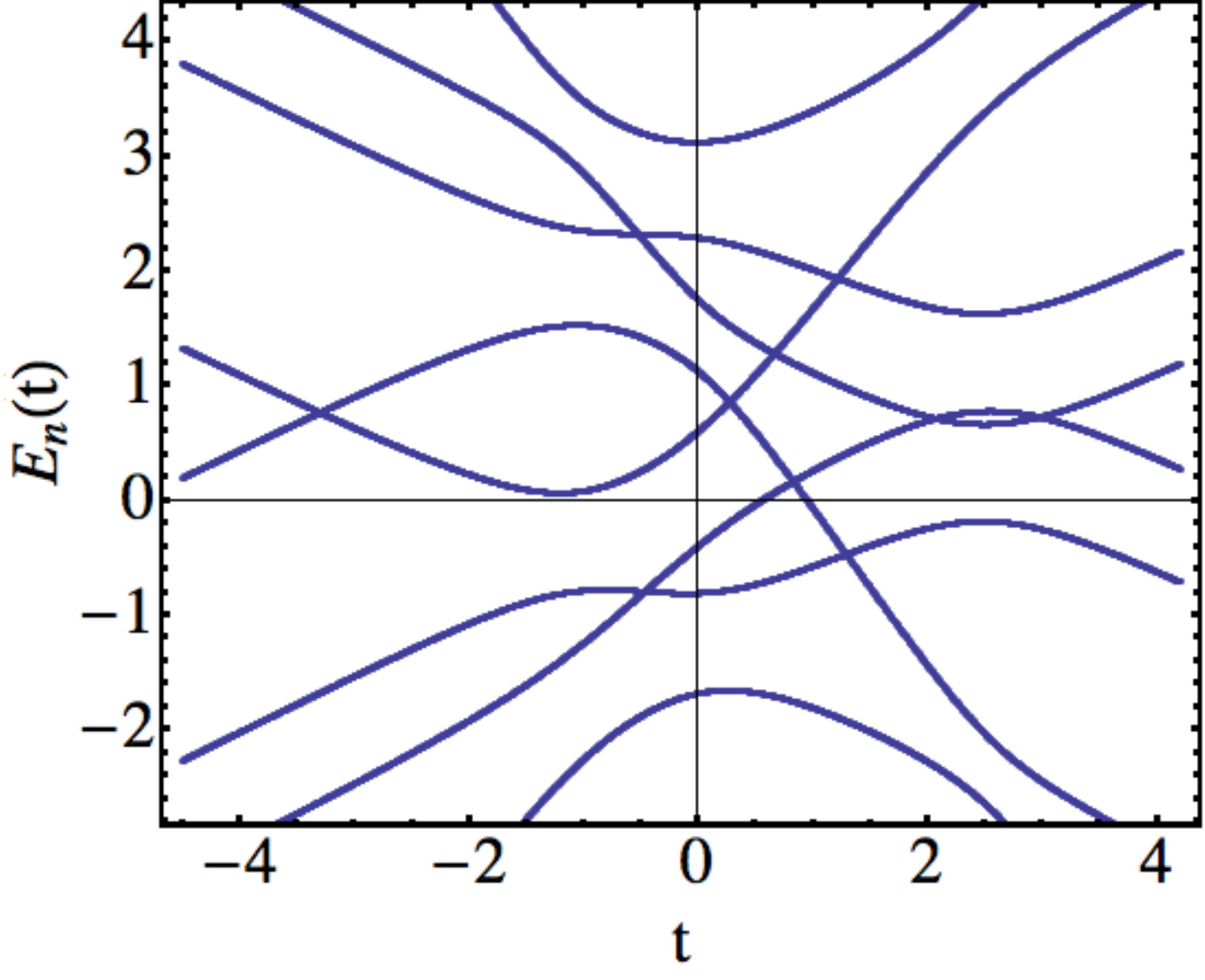}}
\hspace{-2mm}\vspace{-4mm}   
\caption{ (Color online) Eigenvalues of the Hamiltonian (\ref{ham-8}) shown as functions of time (eigenenergy levels). Parameters are: $N_B=1$, $\epsilon_1=2.4$, $\epsilon_2=0$, $\epsilon_3=-1.$, $g=0.35$.
For convenience, equal terms are added to the main diagonal of the Hamiltonian: $\hat{H}\rightarrow \hat{H}-2.5t \hat{1}$.}
\label{a-levels}
\end{figure}

The structure of diabatic levels of the model (\ref{ham-8}) is shown in Fig.~\ref{fig-h8}.  By examining all possible semiclassical trajectories, one can conclude that condition (i) is trivially satisfied because there is always only a single semiclassical path connecting any pair of two states, i.e., there is no interference of different trajectories connecting the initial and final states. Therefore, the dynamic phase factorizes and cancels upon taking the absolute value of the transition amplitude. 

In order to test condition (ii), we diagonalized the Hamiltonian (\ref{ham-8}) and obtained its eigenvalues as functions of time, as shown in Fig.~\ref{a-levels} for $N_B=1$ and $g=0.35$. At such couplings, these functions are already quite distorted from the graph of diabatic levels in Fig.~\ref{fig-h8}. However, one can see visually that Fig.~\ref{a-levels} has ten exact crossing points, which is just the same as the number of diabatic level crossings in Fig.~\ref{fig-h8} that are not marked by direct pairwise couplings. 

Certainly, a Hamiltonian eigenvalue level crossing that is visually looking exact  may hide an invisible mini-gap due to higher order processes. In order to check that this is not the case, we developed a numerical algorithm to estimate the minimal distance between  a pair of eigenenergy levels. According to this algorithm, we increase time near a crossing point in small steps $\delta t$ and trace changes in the difference of eigenenergies, $|E_n-E_{n+1}|$, with nearby indexes. 
We keep going until this difference stops decreasing. Then we make one step back and decrease the size of the step by two orders of magnitude and then repeat the process. After several iterations, we could estimate the minimal difference  $|E_n-E_{n+1}|$ for visually exact crossings to be  less than $10^{-14}$ at parameters chosen in Fig.~\ref{a-levels}. This leaves no doubt that the crossing points in Fig.~\ref{a-levels} are truly exact. Therefore, condition (ii) is satisfied. 

 We also  observed the phenomenon that has not appeared in previous studies of integrable LZ-systems with conditions (i)-(ii) satisfied. At increasing values of coupling parameter $g$, some of the pairs of exact crossing points may approach each other and, eventually, annihilate with each other. For example, by increasing $g$ for the model (\ref{ham-8}) further, only six out of ten exact crossing points survive. It seems, however, that this fact does not break integrability of the model. Consequently, condition (ii) should be reformulated by requiring only that {\it at sufficiently small but finite values of couplings} diabatic  level crossings without direct transitions should give rise to exact crossings of eigenenergy levels. 
 \begin{figure}
\scalebox{0.32}[0.32]{\includegraphics{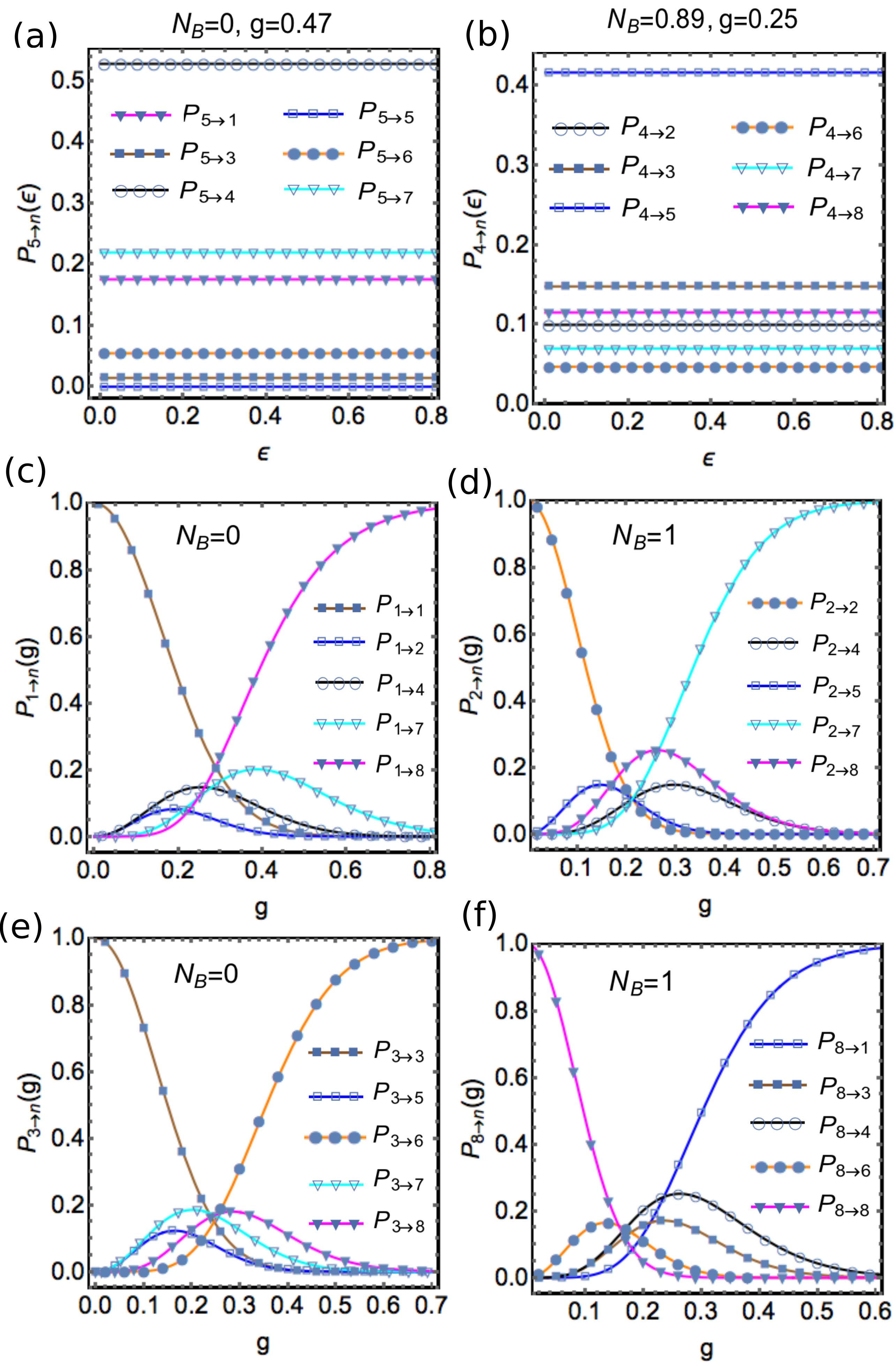}}
\hspace{-2mm}\vspace{-4mm}   
\caption{ (Color online) Comparison of numerical calculations of transition probabilities with semiclassical solution (\ref{prob-8}). All discrete points correspond to results of direct numerical solution of the Schr\"odinger equation with the Hamiltonian (\ref{ham-8}) starting at $t=-1000$ and changing time up to $t=1000$ with a time step size $dt=0.0001$. Solid lines are theoretical predictions of Eq.~(\ref{prob-8}).  In (a) and (b), independence of transition probabilities of energy level splittings is tested for $\epsilon_1=0.5 \epsilon, \epsilon_2=0, \epsilon_3=-\epsilon$. Other parameters are explained in figures. In (c-f), transition probabilities are shown as functions of coupling constant $g$ with different initial conditions at parameter values $\epsilon_1=0.5, \epsilon_2=0, \epsilon_3=-1$. For better visibility, not all possible transition probabilities are shown at given initial conditions.}
\label{numerics}
\end{figure}

\subsection{Solution of Eight-State Model}
Consider the unique allowed semiclassical path from the level-6 to the level-5 that we mark by blue arrows in Fig.~\ref{fig-h8}. 
Since LZ integrability conditions (i)-(ii) are satisfied for the full model (\ref{ham-8}), the exact value of the probability of the transition from the level-6 to the level-5 
is the product of all pairwise LZ transition probabilities encountered during passing or turning at level intersections with nonzero couplings along the semiclassical path. In this case, for example,
$
P_{56}=p_2q_3^2.
$  
Similarly, we can read  probabilities of transitions between any pair of states:
\begin{widetext}
\beq
\hat{P} = \left( 
\begin{array}{cccccccc}
p_1^3          & q_1p_1^2    & p_2q_1p_1 & p_2^2q_1     & q_2q_1p_1 & q_2p_2q_1    & p_3q_2q_1   & q_3q_2q_1 \\
q_1p_2^2   & p_1 p_2^2   & q_2^2 p_2   & q_2^2p_3     & p_2^2q_2    & p_2p_3 q_2   & q_3^2q_2     & p_3q_3q_2\\
p_1q_1p_2 & q_1^2 p_2  & p_2^2 p_1    & p_2q_2^2     & q_2p_1p_2 & q_2^3              &p_3p_2q_2   & q_3p_2q_2\\
p_1^2q_1    &q_1^2 p_1   &p_2q_1^2      & p_2^2 p_1    & q_2q_1^2   & q_2 p_2 p_1   & p_3q_2p_1  & q_3q_2p_1\\
q_1q_2p_3  &q_1q_2p_3 &q_2p_2p_3   & q_2q_3^2     & p_2^2p_3    & p_2 q_3^2      & q_3^2p_3    & p_3^2q_3\\
q_1p_2q_2  &p_1p_2q_2 &q_2^3             & q_2 p_3 p_2 & p_2q_2^2   & p_2^2p_3       &q_3^2p_2     & p_3q_3p_2\\
p_1q_1q_2  &q_1^2q_2    & p_2p_1q_2  & p_2^2 q_2     & q_2^2p_1   & q_2^2p_2       &p_2^2p_3     & q_3p_2^2\\
q_1q_2q_3&p_1 q_2q_3   & q_2p_2q_3  & q_2 q_3 p_3 & p_2^2q_3   & p_2q_3p_3    & q_3p_3^2    & p_3^3
\end{array}
\right).
\label{prob-8}
\eeq
\end{widetext}

 \subsection{Numerical Check}
In order to test our predictions we solved the Schr\"odinger equation for evolution with the Hamiltonian  (\ref{ham-8}) numerically.
We used the algorithm that was described in the supplemental material for Ref.~\cite{sinitsyn-13prl}. According to it, quantum evolution  is simulated from a large negative time and proceeds to a large positive time with very small time steps of a size $dt$. At each time step, the state vector is updated with an evolution operator, which is approximated by
 \beq
 \hat{U}(t,dt) = \left( \hat{1} - i\hat{H}(t) dt/2 \right) \left( \hat{1} + i\hat{H}(t) dt/2 \right)^{-1},
 \eeq
 where $\hat{H}(t)$ is the time-dependent matrix Hamiltonian (\ref{ham-8}), and $\hat{1}$ is the unit matrix. Operator $\hat{U}$(t) is unitary, and up to the order of $o(dt^2)$, it coincides with the true evolution operator, $\hat{\cal T} e^{-i\int_t^{t+dt} \hat{H}(\tau) d\tau}$, where $\hat{\cal T}$ is the time-ordering operator.

 Figure~\ref{numerics} shows excellent agreement between numerical results (discrete points) and theoretical predictions (solid curves) for the $8\times 8$ Hamiltonian (\ref{ham-8}) at {\it all} considered parameter values. 
 One important theoretical prediction of the solution (\ref{prob-8}) is that  all transition probabilities are independent of parameters $\epsilon_i$, as far as we preserve their relative order $\epsilon_1>\epsilon_2>\ldots>\epsilon_{N_s}$. 
In Figs.~\ref{numerics}(a) and (b), we test this prediction by varying these parameters. Importantly,  when they are close to zero, i.e.,  become much smaller than the coupling constants $g_i$, the independent crossing approximation cannot be justified, however, numerical solution still does not show any sign of deviations from  analytical predictions of Eq.~(\ref{prob-8}). 
  Figures~\ref{numerics}(c-f) show transition probabilities as functions of the coupling $g$ for different $N_B$ and different initial state vectors. Again, perfect agreement of  numerical results with Eq.~(\ref{prob-8}) at arbitrary values of parameters  confirms our expectation that Eq.~(\ref{prob-8}) provides, actually, the exact nonperturbative solution of the multistate LZ problem (\ref{ham-8}). 
We also tested solution (\ref{prob-8}) at ``unphysical"  non-integer values of parameter $N_B$  (Fig.~\ref{a11-levels}(b)) and found no difference between the theory and numerical results. In this sense applicability of our solution extends beyond the model (\ref{ham2}).

 The described algorithm is easy-to-realize and it is highly effective when the number of spins is small, $N_s<4$. In such cases, it produces estimates of transition probabilities with accuracy of  two significant digits  for twenty different parameter values in several minutes. 
However, in addition to exponentially growing number of equations at higher values of $N_s$, slow saturation of the solution at large time $t$ and strong oscillations in time require increasingly smaller sizes of time-steps and a larger full time interval for simulations, making such simulations  unreasonably long to persue. It is likely, however, that this numerical approach can be improved by employing interpolation methods that can estimate the solution saturation value by tracing decay of the oscillation envelope, thus reducing the simulation time interval considerably.    
 
 Finally, we note that when a single spin-photon coupling $g$ in (\ref{ham1}) was replaced in our studies by a set of different parameters, $\gamma_i\ne \gamma_j$ for $i\ne j$, that described couplings of different spins to the photon mode, we found that integrability condition (ii) was not satisfied: Minigaps opened up at intersections of diabatic levels without direct couplings. In this case, our numerical studies showed substantial deviations (not shown) from analytical predictions, such as Eqs.~(\ref{bow-tie-prob}) and~(\ref{prob-8}).  These tests provided additional evidence that the agreement of semiclassical analytical predictions and numerical results are not typical for the considered problem and parameter values, except when semiclassical predictions become for some reason exact.

 \section{Higher Dimensional Sectors}
\begin{figure}
\scalebox{0.3}[0.3]{\includegraphics{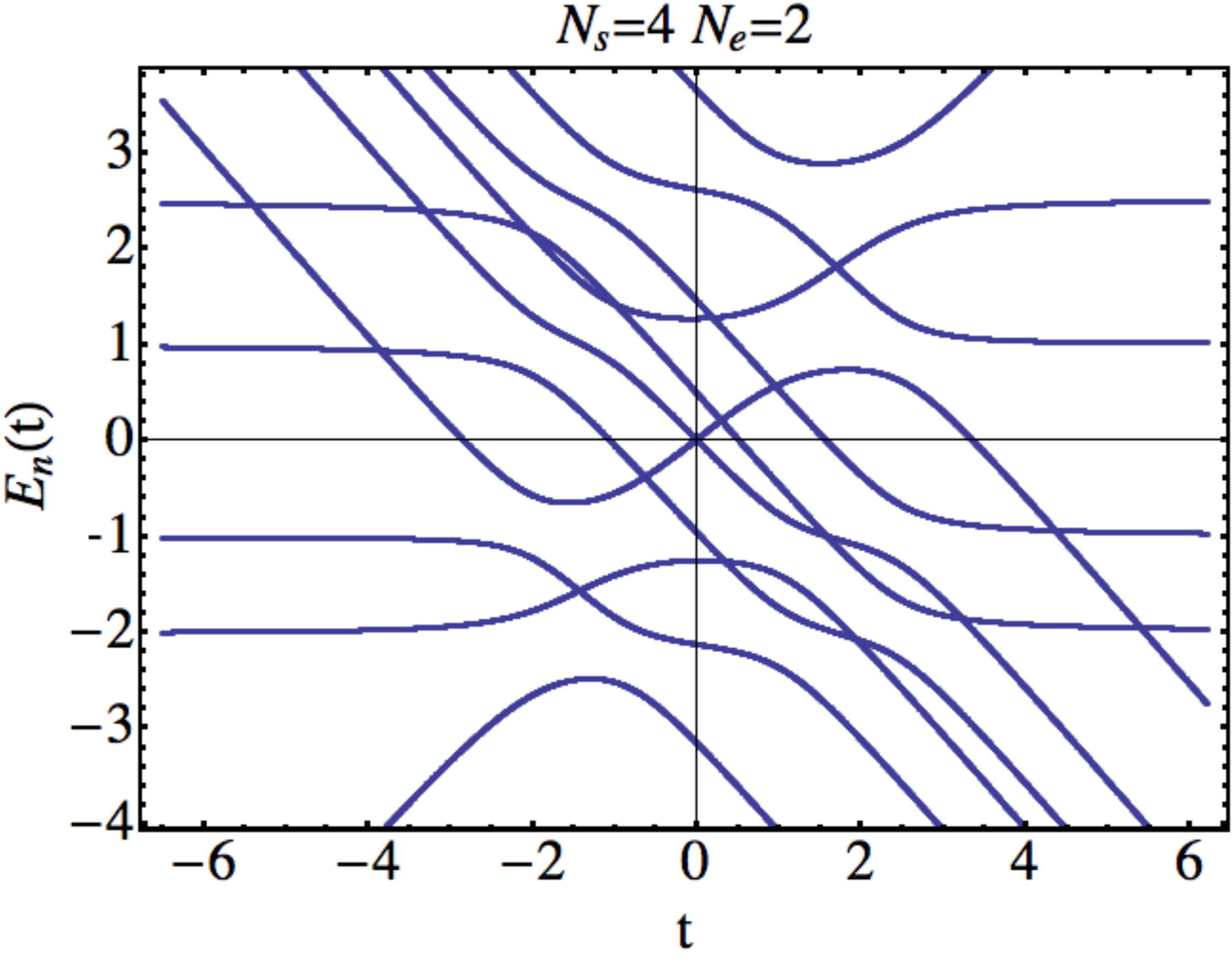}}
\hspace{-2mm}\vspace{-4mm}   
\caption{ (Color online) Eigenenergy levels of the Hamiltonian (\ref{ham2}) in the sector with $N_s=4$ spins and $N_e=2$ excitations, which corresponds to totally eleven interacting states. Parameters are:  $\epsilon_1=2.5$, $\epsilon_2=1$, $\epsilon_3=-1.$, $\epsilon_4=-2.$; $g=0.32$.
For convenience, equal terms are added to the main diagonal of the Hamiltonian: $\hat{H}\rightarrow \hat{H}-3t \hat{1}$.}
\label{a11-levels}
\end{figure}

  Explicit solution development for cases with ever higher number of spins is an increasingly tedious task.  We checked numerically that our observations for $N_s=3$ work  equally well at $N_s=4$. For example, Fig.~\ref{a11-levels} shows eigenenergy levels in the eleven-dimensional sector with $N_e=2$. As expected, there are plenty of exact crossing points to match all crossings of diabatic levels without direct couplings.  
The case with $N_s=4$ generally leads to the sector dimensionality $N=16$ (at $N_e\ge N_s$), which would not be illuminating to develop in this article fully.
We will show in the following section that the degenerate limit, $\epsilon_i=0$, leads to a much smaller subspace dimensionality that we will be able to test numerically up to $N_s\sim 20$.  
 Here, instead, we would like to examine already obtained results and summarize the common properties that we believe are valid for any higher dimensional non-degenerate case.

 Let us compare solution (\ref{prob-8}) for the case $N_s=3$ with (\ref{bow-tie-prob}) for $N_s=2$. We can also extend this set by the case $N_s=1$ that corresponds just to a two-state LZ model with $P_{11}=P_{22}=p_1$ and $P_{12}=P_{21}=q_1$. The following properties appear to be common for all three cases:  
 
(a) There is an algorithm to obtain an explicit  expression for all transition probabilities by drawing the time-energy diagram for diabatic energy levels, such as in Figs.~\ref{do-levels}-\ref{fig-h8}. There is then a unique path connecting any two diabatic states at $t\rightarrow \pm \infty$, such as the path from level-6 to level-5 in Fig.~\ref{fig-h8}, which satisfies conditions of causality and the assumption that diabatic levels can be switched only at level crossings with a nonzero direct coupling between corresponding diabatic states. After identifying such a path, the transition probability of the full trajectory is given by the product of all passing and turning LZ probabilities at encountered level intersections along this path.  
 
(b) A direct consequence of (a) is that we can write down the general formula for all elements along the main diagonal of the transition probability matrix. These elements correspond to the probabilities of events that the system remains at the initial diabatic state after the evolution. Since there is always  a unique semiclassical path for any transition in the model (\ref{ham2}), we can identify such a path in the given case with the path at which the system never turns to other diabatic levels. 
If the initial level has the number  $m$ of spins ``down", then there are $m$ diabatic states with indexes $i$ such that $M(i)=m-1$ that  are directly coupled to the initial state with coupling $g_m$, and there are $(N_s-m)$ states with $M(i)=m+1$ coupled to the initial diabatic state  with coupling $g_{m+1}$. Hence:
\beq
P_{nn}=(p_{M(n)})^{M(n)} (p_{M(n)+1})^{N_s-M(n)},
\label{diag-prob1}
\eeq
 where function $M(n)$ is defined after Eq.~(\ref{pars1}).
 
 (c) At $N_e \ge N_s$, all elements of the transition probability matrix are products of exactly $N_s$ factors $p_i$ or $q_i$, $i=1,\ldots N_s$. Expressions, in terms of $p_i$ and $q_i$ factors, for transition probabilities from any given level to all other states  are  different from each other. 
 
 (d) The transition probability matrix, at $N_e \ge N_s$, has a symmetry,  namely, let us consider elements of this matrix as functions of $p_i$ and $q_i$:
 $P_{nn'} =P_{nn'}(p_1,\ldots,p_{N_s}, q_1, \ldots q_{N_s})$, then
  \begin{eqnarray}
 \nonumber P_{nn'}(p_1,\ldots,p_{N_s}, q_1, \ldots, q_{N_s}) = \\
 P_{N-n+1,N-n'+1} (p_{N_s},\ldots, p_{1}, q_{N_s}, \ldots, q_{1}). 
 \label{prob-sym1}
 \end{eqnarray}
 For example, if we know a compact expression for transition probabilities from the state $n=1$, i.e., with a minimal number of initial bosons, to any other state, then we can use the symmetry (\ref{prob-sym1}) to find a similar expression for transition probabilities from the state with index $N$, i.e. with a maximal number of initial bosons, in the same invariant sector of the model (\ref{ham2}) at $N_e \ge N_s$ .
 
 (e) While writing a general closed expression that describes all possible transition probabilities is an interesting combinatorial problem that we will leave open, it is possible to guess what are transition probabilities for the  physically most important case when the initial level has index $n'=1$, i.e., when the evolution starts with the state having a minimal number of photons in the cavity mode (all spins up). By comparing cases with $N_s=1,2,3$, we introduce a conjecture that, for arbitrary $N_s$ value, the probability of a transition from level $n'=1$ to level $n$ has the form
 \beq
P_{n1}= \left(\prod_{k=1}^{M(n)} q_k \right) \prod_{r=1}^{M(n)+1} p_r^{i_r}, 
\label{conj-1}
\eeq 
 where
\beq
 \quad \sum_{r=1}^{M(n)+1} i_r = N_s-M(n),
\label{conj-2}
\eeq
and where $i_r$ are nonnegative integer numbers. Their zero values are allowed.
 Equation (\ref{conj-2}) does not specify the values of $i_r$ completely but this is not needed if we are interested only in the probability to find a specific number of flipped spins, $m$, at the end of the evolution because the number of all possible combinations of $i_r$ for all levels $n$ with $M(n)=m$ is equal to the number $N_s!/m!(N_s-m)!$ of all  levels with polarization $m$. Hence, all possible values of $i_r$ contribute to the probability to find polarization $m$ after the evolution. 
 One can prove this by noticing that 
 the number of possibilities to organize products of $m+1$ probability parameters $p_1,\ldots,p_{m+1}$ into all possible products made of $N_s-m$ of them, including terms of higher powers of the same probability parameter, can be mapped to the problem of allocating $m+1$ bosons into $N-m$ different discrete states, which results in $N_s!/m!(N_s-m)!$ different possibilities. 

Let $P_{1\rightarrow m}$ be the sum of  $N_s!/m!(N_s-m)!$ probabilities of transitions from level-1 to all states with $M(n)=m$. Then, after using the property (c), we find
 \beq
P_{1\rightarrow m} = \left(\prod_{k=1}^{m} q_k \right) \sum \limits_{i_1,\ldots, i_{m+1}=0}^{N_s-m} \delta_{i_1+\ldots +i_{m+1}, N_s-m} \prod_{r=1}^{m+1} p_r^{i_r} .
\label{conj-3}
\eeq  
 A special case is the probability of the transition from level-1 to level-$N$, i.e., from the state with all spins fully polarized ``up" into the state with fully polarized ``down" spins. In such a case 
 \beq
P_{N1} =P_{1N} = \prod_{k=1}^{N_s} q_k .
\label{conj-4}
\eeq   
 

 \section{Single spin coupled to photon mode}
In this section, we explore the degenerate case: $\epsilon_i=0$ for all $i\in 1,\ldots,N_s$. In the Schr\"odinger equation for amplitudes, any multiplet with $m$ spins ``up" leads to $N_s!/m!(N_s-m)!$ equations that collapse only to a single equation for the symmetric combination of amplitudes, while equations for other orthogonal state amplitudes in this multiplet completely decouple.  
 After decoupling non-symmetric states, the remaining set of equations is equivalent to quantum mechanical evolution with the following Hamiltonian:
\beq
\hat{H}_d = t \hat{a}^{\dagger} \hat{a} +g\left(\hat{a}^{\dagger} \hat{S}^{-}+\hat{a} \hat{S}^+ \right),
\label{ham4}
\eeq
where $S^{\pm}$ are spin raising/lowering operators for a spin size $S=N_s/2$. This time-dependent spin-boson model has attracted lots of attention recently due to applications in the theory of dynamic transitions through a quantum critical point and applications to experiments on molecular condensate creation by a passage of an ultracold atomic gas through a Feshbach resonance. A number of approximate methods have been used to study transition probabilities in the model (\ref{ham4}), including Keldysh technique and adiabatic approximation \cite{gurarie-LZ,painleve-LZ}. It is also known that some of the simpler solvable multistate LZ models correspond to specific limits of the model (\ref{ham4}) \cite{abanov-LZ,chain}.  
Therefore, behavior of the model (\ref{ham4}) is considered well understood, at least in the limit of a large spin ($S\gg 1$). Exact solution can, however, shed light on some cases that are hard to explore with approximate techniques, such as moderate spin sizes and probabilities of rare events.

Here, we will use the same notation, as in previous section, to mark different diabatic states of the model (\ref{ham4}), namely, the  index $n\in (1,\ldots, 2S+1)$ will mark all diabatic levels starting from the lowest number of initial bosons. 
Spin polarization in the model (\ref{ham4}) is related to the index $n$ as 
$
S_z=S-n+1.
$ 

Let us first  work out the case with $S=3/2$ in detail. The Hamiltonian (\ref{ham4}) then describes the sector with four states that are obtained from the eight-state sector (\ref{ham-8}) for the nondegenerate model in the limit $\epsilon_1=\epsilon_2=\epsilon_3=0$. Let $ |1\ra,\ldots, |8\ra $ be the  eight diabatic states in the nondegenerate case. In the degenerate limit, diabatic levels with indexes $n=2,3,4$ become degenerate with each other, as well as levels with indexes $n=5,6,7$. As in the case of the four-state bow-tie model, let us change the diabatic basis and introduce new states:
\begin{eqnarray}
\label{plus2}
 |u_+\ra &\equiv& \frac{1}{\sqrt{3}} \left(|2\ra +|3\ra +|4\ra \right),\\ 
\nonumber |u_1\ra &\equiv& \frac{1}{\sqrt{2}} \left(|2\ra -|4\ra \right), \\
\nonumber |u_2\ra &\equiv& \frac{1}{\sqrt{6}} \left(|2\ra -2|3\ra +|4\ra \right),\\
\label{plus3}
|v_+\ra &\equiv& \frac{1}{\sqrt{3}} \left(|5\ra +|6\ra +|7\ra \right),\\
\nonumber |v_1\ra &\equiv& \frac{1}{\sqrt{2}} \left(|5\ra -|7\ra \right), \\
\nonumber |v_2\ra &\equiv& \frac{1}{\sqrt{6}} \left(|5\ra -2|6\ra +|7\ra \right).
\end{eqnarray} 
In the basis of states $ |1\ra,\, |u_+\ra,\, |v_+\ra,\, |8\ra $, the effective Hamiltonian of the degenerate limit of the model (\ref{ham-8})  reads
\begin{equation}
\hat{H}_d^{(4)}=\left( 
\begin{array}{cccc}
0 & \sqrt{3}g_1 & 0 &0 \\
\sqrt{3}g_1 & t & 2g_2 &0 \\
0 & 2g_2 & 2t & \sqrt{3}g_3 \\
0 & 0& \sqrt{3}g_3 & 3t  
\end{array}
\right).
\label{ham-42}
\end{equation}
One can check by using the spin operator algebra and the definition of parameters (\ref{ppp}) that the Hamiltonian (\ref{ham-42}), indeed, corresponds to the sector $S=3/2$ of the Hamiltonian (\ref{ham4}). 

 \begin{figure}
\scalebox{0.32}[0.32]{\includegraphics{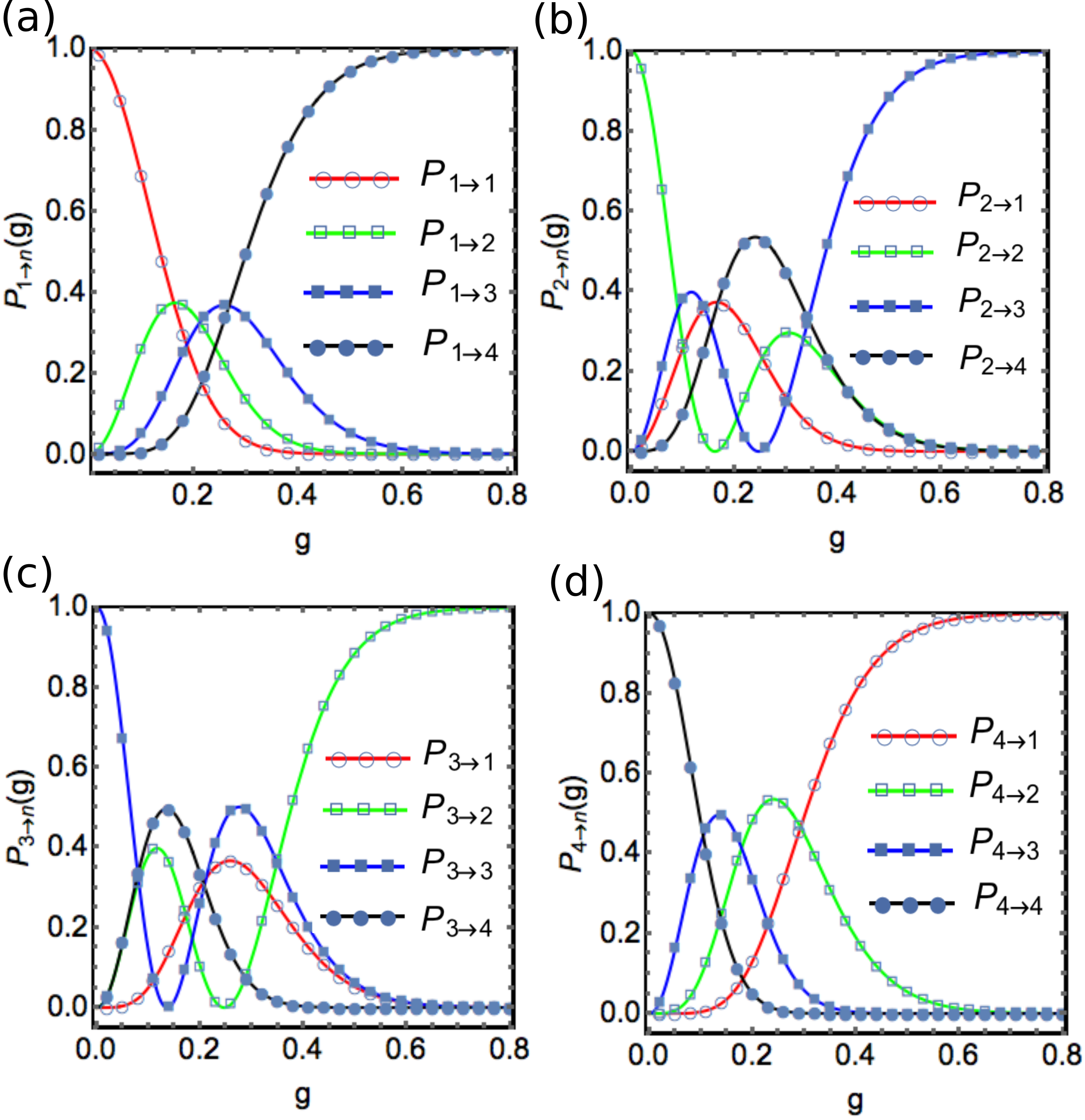}}
\hspace{-2mm}\vspace{-4mm}   
\caption{ (Color online) Transition probabilities for the degenerate case with total spin $S=3/2$. Different sub-figures correspond to different initial conditions. Discrete points correspond to results of numerical simulations of the evolution with the Hamiltonian (\ref{ham-42}) from $t=-1000$ to $t=1000$ with a time step size $dt=0.0001$. Solid lines are theoretical predictions of Eq.~(\ref{prob44}). In all cases, $N_B=1$.  }
\label{LZ4}
\end{figure}
Our steps to derive transition probabilities in the model (\ref{ham-42}) will be analogous to derivation of Eq.~(\ref{bow-tie-short}) for the case of the three state bow-tie model, which also corresponds to the sector  $S=1$ of the model (\ref{ham4}). Thus, transition probabilities to/from the levels with extremal slopes ($S_z=\pm 3/2$) from/to states $|u_+\ra$ or $|v_+\ra$ are given by the sum of transition probabilities to all corresponding  diabatic states  in the model (\ref{ham-8}) that become degenerate. For example, in order to find the transition probability from level-2 to level-1 of the model (\ref{ham-42}) one should take the sum of transition probabilities from levels-2,3,4 to level-1 of the model (\ref{ham-8}):
$$
P^{(4)}_{12}=q_1(p_2^2+p_1p_2+p_1^2).
$$
In the case of transitions between $ |u_+\ra $ and $ |v_+\ra$, we have to include information about probabilities to stay/leave to/from the decoupled states. The state $|u_1\ra$ is coupled only to the state $|v_1\ra$ with coupling $g_2$, and $|u_2\ra$ is coupled to $|v_2\ra$ also with coupling $g_2$. Corresponding transition probabilities between those pairs of states are described by the standard 2-state LZ transition probability:
\begin{eqnarray}
\label{dec1}
P_{u_1,u_1}&=&P_{v_1,v_1} =P_{u_2,u_2}= P_{v_2,v_2}=p_2, \\
\label{dec2}
 P_{v_1,u_1}&=&P_{v_2,u_2}=P_{u_1,v_1}=P_{u_2,v_2} = q_2.
\end{eqnarray}
By analogy with  derivation of the probability to stay in the second level in the degenerate three-state bow-tie model that we described at the end of section 2,  we equate the sums of transition probabilities between sets of ``becoming-degenerate" levels in the basis 
$ |1\ra,\ldots, |8\ra $  of the model  (\ref{ham-8}) and the sums of corresponding transition probabilities in the basis (\ref{plus2})-(\ref{plus3}):
$$
P_{22}^{(4)} +P_{u_1,u_1}+P_{u_2,u_2} = \sum \limits_{i,j=2,3,4} P_{ij},
$$
where $P_{ij}$ are elements of the eight-state probability matrix (\ref{prob-8}). Similarly, 
$$
P_{32}^{(4)} +P_{v_1,u_1}+P_{v_2,u_2} = \sum \limits_{i=2,3,4} \sum \limits_{j=5,6,7} P_{ji}.
$$
We can now summarize all such results in a single probability matrix for the model (\ref{ham-42}), which we test numerically in Fig.~\ref{LZ4}:
\begin{widetext}
\begin{equation}
\hat{P}^{(4)}=\left( 
\begin{array}{llll}
 p_1^3 &  q_1(p_1^2+p_1p_2 +p_2^2) & q_1q_2(p_1+p_2+p_3) & q_1q_2q_3 \\
 q_1(p_1^2+p_1p_2 +p_2^2)& A & C & q_3q_2(p_3+p_2+p_1) \\
 q_1q_2(p_1+p_2+p_3)& C &B &q_3(p_3^2+p_2p_3+p_2^2) \\
q_1q_2q_3& q_3q_2(p_3+p_2+p_1)&q_3(p_3^2+p_2p_3+p_2^2)& p_3^3 \\
\end{array}
\right),
\label{prob44}
\end{equation}
where 
$$
C=q_2(p_2^2+2p_2(p_3+p_1) +p_1p_3+q_1^2+q_2^2+q_3^2-2),$$
$$A=3p_2^2p_1+p_1q_1^2+p_3q_2^2+2p_2(q_1^2+q_2^2-1),
$$ 
$$
B=3p_2^2p_3+p_1q_2^2+p_3q_3^2+2p_2(q_2^2+q_3^2-1).
$$
\end{widetext}

Here, we would like to point to one useful symmetry of the transition probability matrix (\ref{prob44}): 
\beq
P^{(4)}_{ij}=P^{(4)}_{ji}.
\label{sym2}
\eeq
This property is, actually, valid for any higher dimensional sector of the model (\ref{ham4}) because, in the matrix form, this model belongs to the class of multistate LZ-chain models, in which only couplings between states with nearest indexes are nonzero. Indeed, for spin $S$, the only independent nonzero matrix elements of the Hamiltonian (\ref{ham4}) are
$
\left( H_d^{(2S+1)} \right)_{nn}=(n-1)t$, and
$$
 \left(H_d^{(2S+1)}\right)_{n,n+1}=g_n\sqrt{S(S+1)-(S-n+1)(S-n)}.
$$ 
The symmetry (\ref{sym2}) was rigorously proved in \cite{lzc-14pra} for any LZ-chain model. 

The worked out examples for $S=1$ and $S=3/2$ demonstrate that there is a simple algorithm to obtain the transition probability matrix for arbitrary $S$. Symmetric combinations of ``becoming-degenerate" diabatic states and extremal slope states  of the model (\ref{ham2}) correspond to the diabatic states of the desired sector of the model (\ref{ham4}). For the latter, transition probabilities become combinations of ones in the nondegenerate model (\ref{ham2}) with $N_s=2S$ and transition probabilities of   the lower-$S$ sectors of the degenerate model.

While there is no doubt that it is possible to automate the process of deriving explicit  expressions for the full transition probability matrices and  fully solve a few higher-dimensional sectors, the number of terms involved is growing very quickly with $S$. Instead, in the rest of this section, we will focus on the physically most interesting case when the system starts with a fully polarized state $S_z=\pm S$, i.e., $n=1$ or $n=2S+1$. 
First, we note that Eq.~(\ref{conj-3}) applies equally to the model (\ref{ham4}), where we can identify parameter   $N_s$ with $2S$ and parameter $m$ with an index $n-1$ of the level of the Hamiltonian (\ref{ham4}) to which the transition  is considered. So, in terms of parameters of the model (\ref{ham4}), we have
 \beq
P_{1\rightarrow n} = \left(\prod_{k=1}^{n-1} q_k \right) \sum \limits_{i_1,\ldots, i_{n}=0}^{2S+1-n} \delta_{i_1+\ldots +i_{n}, 2S+1-n} \prod_{r=1}^{n} p_r^{i_r} .
\label{conj-4}
\eeq 
Moreover, using the symmetry (\ref{prob-sym1}), we can write a formula for the transition from the state number $2S+1$:
 \begin{eqnarray}
  \label{conj-5}
&&P_{2S+1\rightarrow n} = \left(\prod_{k=n}^{2S} q_{k} \right) \times \\
\nonumber  && \sum \limits_{i_{(n-1)},\ldots, i_{2S}=0}^{n-1} \delta_{i_{(n-1)}+\ldots +i_{2S}, n-1} \prod_{r=n-1}^{2S} p_r^{i_r} .
\end{eqnarray}

\begin{figure}
\scalebox{0.24}[0.24]{\includegraphics{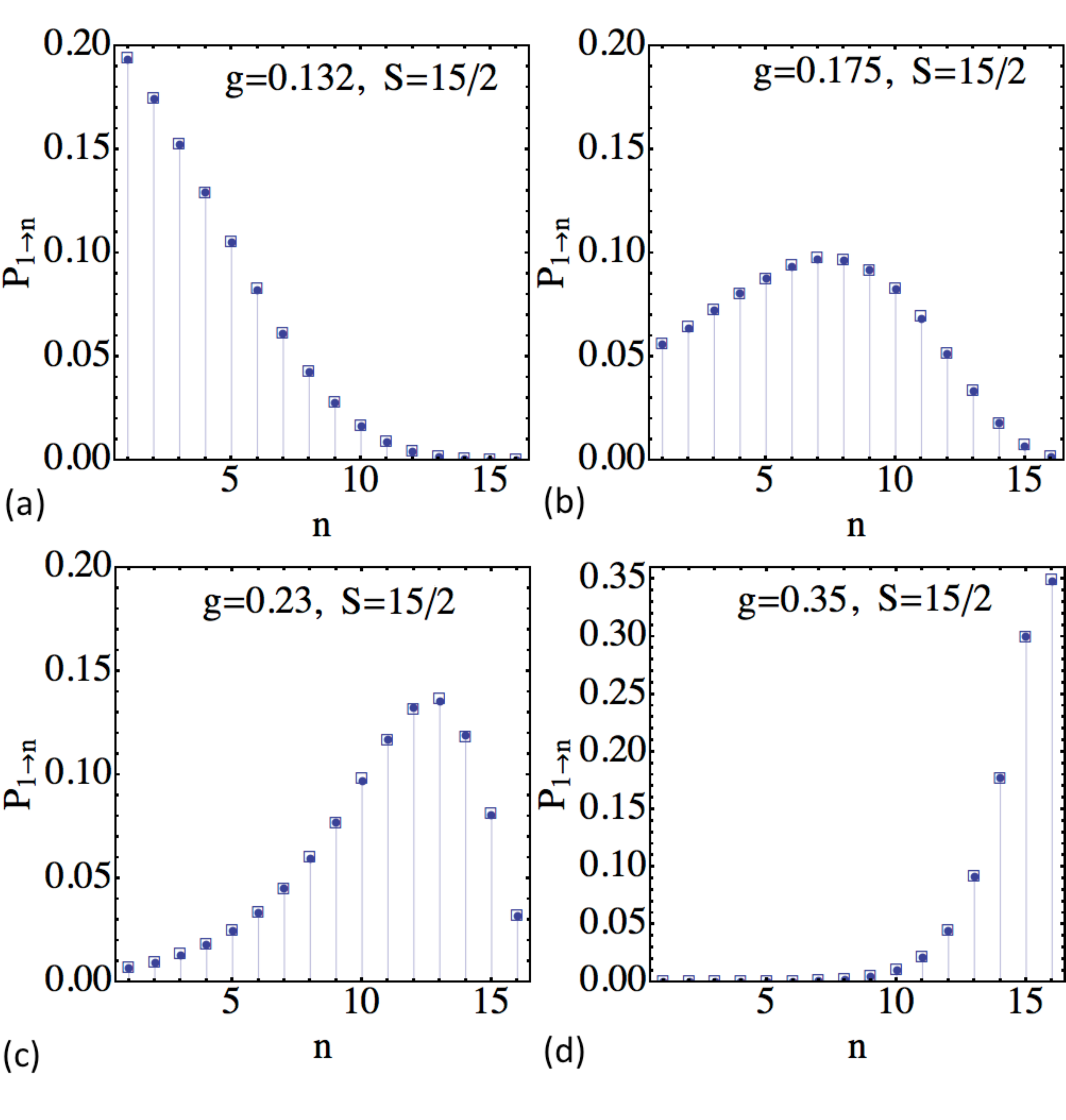}}
\hspace{-2mm}\vspace{-4mm}   
\caption{ (Color online) Transition probabilities $P_{1\rightarrow n} \equiv P^{(2S+1)}_{n1}$ from the state $n=1$ ($S_z=S$) to other states for evolution with the Hamiltonian (\ref{ham4}) in the sector with 16 states ($S=15/2$) at different coupling strengths.  Theoretical predictions (dots on vertical lines) of Eq.~(\ref{conj-4}) are compared to results of direct numerical simulations (empty boxes)  Time evolution is from $t=-2000$ to $t=2000$ with time step $dt=0.00005$. In all cases, $N_B=0$.}
\label{n16-fig}
\end{figure}

\begin{figure}
\scalebox{0.24}[0.24]{\includegraphics{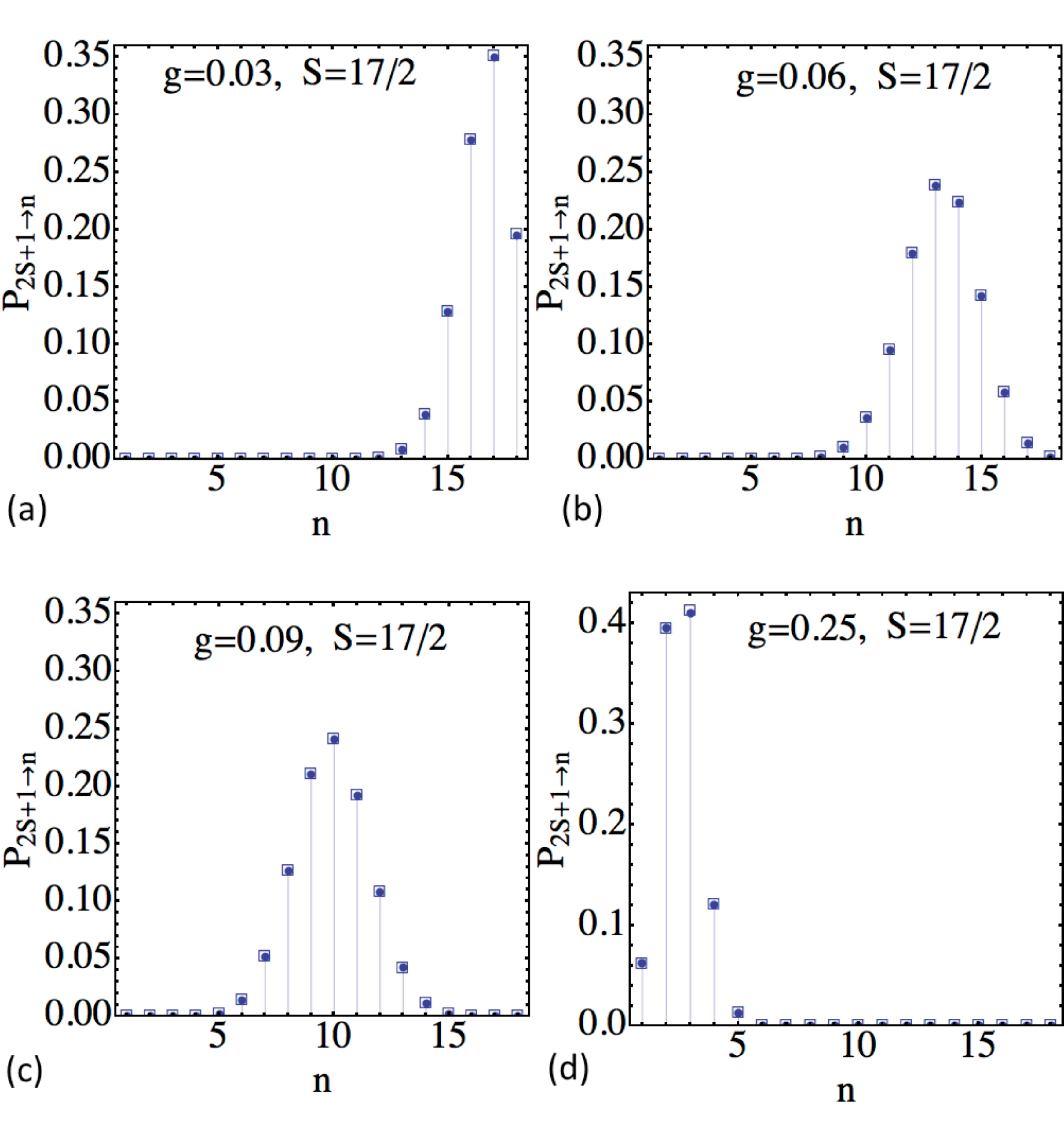}}
\hspace{-2mm}\vspace{-4mm}   
\caption{ (Color online) Comparison of theoretical predictions (dots on vertical lines) of Eq.~(\ref{conj-5}) and numerical simulations (empty boxes) for evolution with the Hamiltonian (\ref{ham4}) in the sector with 18 states ($S=17/2$) at different coupling strengths. Time evolution is from $t=-2000$ to $t=2000$ with time step $dt=0.00005$. In all cases, $N_B=0$. Evolution starts with the fully polarized spin state $S_z=-S$ with $2S$ bosons, which corresponds to $n=2S+1=18$.}
\label{n18-fig}
\end{figure}

In Figs.~\ref{n16-fig} and \ref{n18-fig}, we compare theoretical predictions of Eqs.~(\ref{conj-4})-(\ref{conj-5}) with our results of the direct numerical solution of the Schr\"odinger equation for two different sectors of the model (\ref{ham4}). 
Since the matrix Hamiltonian of the model (\ref{ham4}) has nonzero elements only along the main and the next to the main diagonals, we can accelerate our simulations by applying the leap-frog algorithm to invert matrices, which scales as $N{\rm Log}(N)$ with the size $N$ of the matrix. This allows us to check our theoretical predictions for considerably larger phase space sizes than in previous sections. Therefore,  Figs.~\ref{n16-fig} and \ref{n18-fig}  serve also as a test of our conjectures made about higher dimensional sectors of the model (\ref{ham2}).

Figure~\ref{n16-fig} corresponds to the initially polarized state without bosons. Subfigures (a)-(d) show results at progressively larger values of the coupling parameter $g$. At moderate values of coupling, there is a phase (case (b-c)) at which practically all states have substantially nonzero probabilities to be populated at the end of the evolution. This is the result of the positive feedback induced by the emission of bosons, which leads to amplification of quantum fluctuations. 

Figure~\ref{n18-fig} shows transition probabilities in the opposite process when the spin is initially fully polarized ``down" but there are $2S$ photons that it can absorb. In this case, very small values of $g$ are sufficient to produce substantial transition probabilities, as shown in Fig.~\ref{n18-fig}(a). By increasing $g$, the final distribution has a shape of a localized wavepacket that moves towards the state with zero number of bosons with increasing $g$ (Figs.~\ref{n18-fig}(b-c)). Eventually, at large $g$, the final distribution becomes sharply peaked near the state with $S_z=S$. 

\section{Discussion}
We explored the cavity QED model of spins interacting with a single optical mode in a linearly time-dependent field. This model and its degenerate version (\ref{ham4}) have  already encountered in the literature with diverse applications, from cavity QED to transitions through the Feshbach resonance in molecular condensates \cite{gurarie-LZ,painleve-LZ,abanov-LZ}. Certainly, existence of an exact solution can help to extend the discussion of these applications. For example, the semiclassical limit of the model (\ref{ham4}) corresponds to an explicitly time-dependent classical Hamiltonian dynamics with connections to the Painlev\'e (PII) equation \cite{painleve-LZ}. It would be insightful to explore whether the semiclassical limit of our solution can lead to the exact solution of such a classical system.

Despite a number of fully solvable multistate LZ models has been known, all such models  describe relatively simple situations, such as interactions of many uncoupled to each other levels with a single level \cite{do,bow-tie1}.
Many known solvable cases are, in fact, reducible in the sense that they can be decoupled into a set of independent  Demkov-Osherov, bow-tie, or  two-state LZ models  by applying simple well-characterized symmetry transformations \cite{multiparticle}. Only recently, conditions of integrability, which we discussed in section 3B, have been used by one of us to uncover a few more relatively small-size solvable models \cite{six-LZ,four-LZ}. 
  
Perhaps, the most important achievement of our work is the demonstration that systems with very complex many-body interacting Hamiltonians can be solvable at conditions of arbitrarily strong nonadiabatic driving of model parameters.  Surprisingly, our final solution for the transition probability matrix appears to be relatively simple; it is expressed only via elementary functions. This is in contrast to the time-independent case, whose solution generally can be written only implicitly via the roots of nonlinear algebraic equations. Moreover, analysis of the independent crossing approximation, which we used to obtain transition probability matrices in this article, appears to be even simpler than in
models  \cite{six-LZ,four-LZ} that were previously derived using the same integrability conditions. Namely, in the case of the model (\ref{ham2}), there is no interference between different semiclassical trajectories that connect different initial and final states in the independent crossing approximation. This strongly suggests that the class of fully integrable multistate LZ models can be substantially extended and it is important to continue developing the theory of  quantum integrability of  multistate LZ systems in order to obtain the means to deal with strongly interacting nonstationary problems in quantum mechanics. 

Several questions need to be resolved within this theory. Until now, there is no mathematically rigorous proof of the LZ integrability conditions introduced in  \cite{six-LZ,four-LZ}. Their validity is supported only by considerable amount of numerical tests for models that have been solved by using such conditions and by a number of  models that have been solved analytically by alternative methods. It is unclear whether there are counterexamples or further restrictions on these conditions. They do not provide a direct path to derive new solvable models. Rather they serve as a test that a model should pass in order to be solvable by the semiclassical ansatz.  

The most important clue that LZ integrability conditions provide is that they request the existence of  exact adiabatic energy crossing points that must encounter when one is varying a single model parameter, which is also often the property of quantum integrable systems solvable by the algebraic Bethe ansatz. Recently, the reasons for the presence of such crossing points were questioned, and a new type of a dynamic symmetry based on the existence of nontrivial ``commuting partner Hamiltonians" was suggested to explain it  \cite{com-partner}. Both Demkov-Osherov and bow-tie models have been shown to possess such commuting partners \cite{armen}. This observation may be used to uncover new  candidate LZ models that can be exactly solved. Finally, we would like to mention that the Stokes phenomenon has peculiar properties in multistate LZ systems  \cite{lzc-14pra,sun-jpa15}. Its relation to the LZ-integrability is another peace of the puzzle.


{\it Acknowledgment}. Authors thank Avadh Saxena for useful discussions. The work
was carried out under the auspices of the National Nuclear
Security Administration of the U.S. Department of Energy at Los
Alamos National Laboratory under Contract No. DE-AC52-06NA25396. Authors also thanks the support from the LDRD program at LANL.


\begin{thebibliography}{100}
\bibitem{note-qed} Physically more justified Hamiltonian may also contain terms  of the type $a_k(t)\hat{1}_k$, where $\hat{1}_k$ is the unit operator in the space the $k$th spin, and $a_k(t)$  is an arbitrary time-dependent parameter.
 Such terms split relative energies of different spins, however, being commuting with each other and with the rest of the Hamiltonian (\ref{ham1}), they can be removed from the Schr\"odinger equation by a trivial gauge transformation: 
 $\psi(t)\rightarrow \exp \left(-i\int_{-\infty}^t dt' \, \sum_k a_k(t')\hat{1}_k \right) \psi(t)$.
\end{thebibliography}
\end{document}